\documentclass[12pt]{iopart}

\usepackage{latexsym}

\newcommand{\del}{\partial}
\newcommand{\beq}{\begin{eqnarray}}
\newcommand{\eeq}{\end{eqnarray}}
\newcommand{\be}{\begin{eqnarray*}}
\newcommand{\ee}{\end{eqnarray*}}
\newcommand{\bk}{{\bf k}}

\newcommand{\bx}{{\bf x}}

\newcommand{\ra}{\rightarrow}

\newcommand{\nn}{\nonumber}

\newcommand{\ex}[1]{\langle\,#1\rangle}
\newcommand{\D}{{\cal D}}

\begin{document}

\title{Resolution of an apparent inconsistency in the electromagnetic Casimir effect}

\author{H. Alnes$^1$, K. Olaussen$^2$, F. Ravndal$^1$ and I.K. Wehus$^1$}

\address{$^1$ Department of Physics, University of Oslo, N-0316 Oslo, Norway.}

\address{$^2$  Department of Physics, NTNU, N-7491 Trondheim, Norway.}

\ead{i.k.wehus@fys.uio.no}

\begin{abstract}
The vacuum expectation value of the electromagnetic energy-momentum tensor between two parallel plates in spacetime dimensions $D > 4$ is calculated in 
the axial gauge. While the pressure between the plates agrees with the global Casimir force, the energy density is divergent at the plates and not 
compatible  with the total energy which follows from the force. However, subtracting the divergent self-energies of the plates, the resulting
energy is finite and consistent with the force. In analogy with the corresponding scalar case for spacetime dimensions $D>2$, the divergent self-energy of 
a single plate can be related to the lack of conformal invariance of the electromagnetic Lagrangian for dimensions $D > 4$.
\end{abstract}

Two parallel, metallic plates separated by the distance $L$ in vacuum, will interact due to the modifications of the quantum fluctuations of the 
electromagnetic field caused by the boundary conditions at the plates. The resulting force was first calculated by Casimir\cite{Casimir} who found it to be 
given by the attractive pressure  $P = - {\pi^2/240L^4}$. 
Using the conformal symmetry of the electromagnetic field in $D=4$ spacetime dimensions, Brown and Maclay\cite{Lowell} later obtained the vacuum 
expectation values of all the components of the electromagnetic energy-momentum tensor
\beq
               T_{\mu\nu} = F_{\mu\alpha}F^\alpha_{\;\;\,\nu}  -
	       \eta_{\mu\nu} {\cal L}    \label{T_EM}
\eeq
where ${\cal L} = -(1/4)F_{\alpha\beta}^2$ is the standard Lagrangian. While these expectation values were constant between the plates, the corresponding 
fluctuations of the separate electric and magnetic fields were found by L\"utken and Ravndal  to be in general non-constant and actually divergent 
as one approaches one of the plates\cite{LR}. These divergences are caused 
by imposing ideal boundary conditions valid for arbitrarily small wavelengths of the field. A physical boundary would only affect fluctuations down to a 
finite wavelength which is expected to result in an increasing, but finite value of the fluctuations near the plates. The quantitative effects of such more 
realistic boundary conditions have been investigated during the last few years but a complete and satisfactory description is still lacking\cite{phys_bc}. 

Casimir forces in spacetimes with dimensions $D > 4$ were first systematically calculated by Ambj\o rn and Wolfram\cite{AW}. For the electromagnetic 
field between two parallel hyperplanes with separation $L$, the attractive pressure was found to be
\beq
         P = - (D-1)(D-2) {\Gamma(D/2) \zeta_R(D)\over (4\pi)^{D/2}L^D}                                        \label{press}
\eeq
where the factor $D-2$ is the number of physical degrees of freedom in the field resulting from gauge invariance. If the energy density between the plates 
is constant, it would just be this pressure divided by the factor $D-1$. This is the case when $D=4$ and it is of interest to see if it holds also 
in the more general case $D>4$.  For this purpose we calculate in the following the separate fluctuations of the electric and magnetic components of the 
field which then allows us to find all the vacuum expectation values of the components of the energy-momentum tensor (\ref{T_EM}).

Today these quantum effects could be of relevance for stacks of parallel branes where the electromagnetic field is replaced by one or more of  the 
abelian Ramond-Ramond fields. Any divergent  energy density would then have serious implications for the stability of such configurations due to the 
resulting large gravitational interactions.

The electromagnetic field tensor $F_{\mu\nu} = \del_\mu A_\nu - \del_\nu A_\mu $ in $D=d+1$ spacetime dimensions has $d$ electric components 
$E_i = F_{0i}$ and $d(d-1)/2$ magnetic components $B_{ij} = F_{ij}$. 
For the geometry under consideration, the simplest and most natural choice of gauge is the axial gauge $n^\mu A_\mu = 0$ where the unit $D$-vector $n^\mu$ 
is normal to the plates. Taking this along the $z$-axis, we thus have $A_z = 0$. The component $A_0$ is no longer a free variable, 
but depends on the others via the Maxwell equation $\del_iF^{i0} = 0$. It gives $A_0 = -\Delta^{-1}\del_i {\dot A}_i$ where the operator 
$\Delta = \del_i^2$. There are thus $D-2$ independent degrees of freedom described by the spatial field components $A_i$ where $i \ne z$. 
The full Lagrangian then follows as
\beq
             L = {1\over 2}\int\!d^dx \left[{\dot A}_i\Big(\delta_{ij} - \del_i\Delta^{-1}\del_j\Big){\dot A}_j
               -  A_i\Big(\del_i\del_j - \delta_{ij}\Delta\Big)A_j\right]                                                   \label{Lagrange}
\eeq
after a few partial integrations and neglecting surface terms.

In order to quantize the system, we must solve the classical wave equation following from the Lagrangian. For this purpose we impose the boundary condition 
$n^\mu F_{\mu\nu} = 0$ at the plates. This is the same as for the MIT quark bag where it had a physical justification\cite{MIT}. Here it is just taken 
for convenience. In the axial gauge it gives $\del_z A_i = 0$ at the plates which is the Neumann boundary condition for each physical field  component 
$A_i(x) = A_i(t;\bx_T,z)$. We then have the general mode expansion 
\beq
              A_i(t;\bx_T,z) = \sqrt{2\over L} \sum_{n=1}^\infty \int\! {d^{d-1}k_T\over (2\pi)^{d-1}} A_{in}(t,\bk_T) e^{i\bk_T\cdot \bx_T} 
                                \cos\left({n\pi z\over L}\right)     \label{mode-ex}
\eeq
which satisfies the wave equation and the boundary conditions. The factor $\sqrt{2/L}$ is a normalization factor. In the mode sum we have dropped a 
term with $n=0$ since it will not contribute to any physical results after regularization.

Quantization can now be done in the standard way. We introduce orthonormal polarization vectors ${\bf e}_\lambda$ normal to the wavevector 
$\bk_T$ and a longitudinal polarization vector ${\bf e}_L$ along this direction. The coordinate components $A_{in}$ of the field are then replaced 
by the polarization components $(A_{\lambda n}, A_{L n})$. After quantization at $t=0$ the transverse components can then be written on the standard form as
\beq
       A_{\lambda n}(\bk_T) = \sqrt{1\over 2\omega_n}\Big[a_{\lambda n}(\bk_T)  + a_{\lambda n}^\dagger(-\bk_T) \Big]              \label{comps}
\eeq
where $\omega_n^2 = \bk_T^2 + k_z^2$ with $k_z = \pi n/L$. The creation and annihilation operators now have the standard commutator
\beq
         [a_{\lambda n}(\bk_T) ,a_{\lambda'n'}(\bk'_T)] = \delta_{\lambda\lambda'}\delta_{nn'}(2\pi)^{d-1}\delta(\bk_T - \bk'_T)          \label{comm}
\eeq
However, the longitudinal component 
\beq
      A_{Ln}(\bk_T) = \sqrt{1\over 2\omega_n}\left({\omega_n\over k_z}\right)\Big[a_{Ln}(\bk_T)  + a_{Ln}^\dagger(-\bk_T) \Big]
\eeq
contains an extra factor when the corresponding creation and annihilation operators have the same canonical commutator (\ref{comm}). The full field
operator (\ref{mode-ex}) is then expressed in terms of these new operators corresponding to definite polarization states.

The field fluctuations between the two plates can now easily be calculated. As a simple example, consider $E_z = -\del_z \Delta^{-1}\del_j{\dot A}_j$. If
we isolate the mode with quantum numbers $(n,\bk_T)$, we find the operator
\beq
    \Delta^{-1}\del_j{\dot A}_j =
\nonumber\\
  \sqrt{2\over L}\sqrt{1\over 2\omega_n}
    \frac{(ik_j)(-i\omega_n)}{\omega^2_n}
    \left[a_{\lambda n}e_{\lambda j} + \left({\omega_n\over k_z}\right)a_{Ln} e_{Lj}\right] e^{i\bk_T\cdot \bx_T}\cos\left({n\pi z\over L}\right) + \mbox{H.c}.
\eeq
acting on the vacuum state. Here we have used that  $\Delta$ gives  $\bk_T^2 + k_z^2 = \omega_n^2$ in momentum space. We see that the transverse modes
will not contribute here since they satisfy the orthogonality condition ${\bf e}_\lambda\cdot\bk_T = 0$. However, for the longitudinal mode we have
instead ${\bf e}_L\cdot\bk_T = k_T$ and it will give a non-zero contribution. The derivative $\del_z$ gives a factor $k_z$ and 
$\cos(n\pi z/L)\ra \sin(n\pi z/L)$. For this mode alone we thus get the fluctuation
\beq
    \ex{E_z^2}|_{n,\bk_T} = {2\over L}{1\over 2\omega_n} k_T^2\sin^2{n\pi z\over L}  
\eeq
Including all the modes, we thus have for the full fluctuation of this electric field component
\beq
     \ex{E_z^2} &=& {1\over L}\sum_{n=1}^\infty \int\! {d^{d-1}k_T\over (2\pi)^{d-1}}\left[\omega_n - {k_z^2\over\omega_n} \right]\sin^2{n\pi z\over L} 
\eeq
when we write $k_T^2 = \omega_n^2 - k_z^2$. For the other components we similarly find
\beq
    \ex{E_i^2} &=& {1\over L}\sum_{n=1}^\infty \int\! {d^{d-1}k_T\over (2\pi)^{d-1}}\left[\omega_n(d-2) +{k_z^2\over\omega_n} \right]\cos^2{n\pi z\over L}
\eeq
where there is an implied sum over the transverse index $i$. The magnetic field fluctuations can be obtained the same way and become
\beq
  \ex{B_{iz}^2} &=& {1\over L}\sum_{n=1}^\infty \int\! {d^{d-1}k_T\over (2\pi)^{d-1}}\left[\omega_n + {k_z^2\over\omega_n}(d-2) \right]\sin^2{n\pi z\over L} \\
 \ex{B_{i<j}^2} &=& {1\over L}\sum_{n=1}^\infty \int\! {d^{d-1}k_T\over (2\pi)^{d-1}}\left[\omega_n(d-2) -{k_z^2\over\omega_n}(d-2) \right]\cos^2{n\pi z\over L}
\eeq
when we again sum over the indices $i$ and $j$. We have also confirmed
these results by 
performing the same calculations in Coulomb gauge instead of axial gauge.

Using now a combination of dimensional and zeta-function
regularization as previously used  when $D=4$\cite{TR}, we can write
the  result on the form
\beq
{1\over 2L}\sum_{n=1}^\infty \int\! {d^{d-1}k_T\over (2\pi)^{d-1}}
\left(1 \pm \cos{2n\pi z\over L}\right)\left\{
\begin{array}{c}\omega_n\\{k_z^2/\omega_n}\end{array}
\right\}
\nn \\ 
= 
 -{\Gamma(D/2)\over (4\pi)^{D/2}L^D}\left[\zeta_R(D) \pm {1\over 2}f_D(z/L)\right] 
\left\{
\begin{array}{c}1\\D-1\end{array}
\right\}
\eeq
Here $\zeta_R(D)$ is the Riemann zeta-function while $f_D(z/L)$ depends on the distance $z$ from the plates. When the spacetime dimension $D$ is even, it can
be written on the compact form
\beq
        f_D(z/L) = {\pi^D\over \Gamma(D)}\left(-{d\over d\theta}\right)^{D-1} \cot\theta  \hspace{10mm} (D = \mbox{even})          \label{Deven}
\eeq
where $\theta = \pi z/L$. But when $D$ is odd, no such closed expression is easily derived. However, using a different regularization based on the 
corresponding point-split Green's functions, one finds in general\cite{AORW}
\beq
      f_D(z/L)  = \sum_{j=-\infty}^\infty {1\over |j + z/L|^{D}}  = \zeta_H(D,z/L) +  \zeta_H(D,1 - z/L)                                      \label{f_D}
\eeq
where  $\zeta_H(D,z/L)$ is the Hurwitz zeta-function. When  $D$ is even, this can be shown to agree with (\ref{Deven}). 

The regularized fluctuations of the electric field normal to the plates thus become
\beq
          \ex{E_z^2} = {(D-2)\Gamma(D/2)\over (4\pi)^{D/2}L^D}\left[\zeta_R(D) - {1\over 2}f_D(z/L)\right]
\eeq
while for the transverse components we find
\beq
          \ex{E_i^2} = -2{(D-2)\Gamma(D/2)\over (4\pi)^{D/2}L^D}\left[\zeta_R(D) + {1\over 2}f_D(z/L)\right]
\eeq
For the magnetic fluctuations we similarly have
\beq
          \ex{B_{iz}^2} = -(D-2){(D-2)\Gamma(D/2)\over (4\pi)^{D/2}L^D}\left[\zeta_R(D) - {1\over 2}f_D(z/L)\right]
\eeq
and
\beq
          \ex{B_{i<j}^2} = (D-3){(D-2)\Gamma(D/2)\over (4\pi)^{D/2}L^D}\left[\zeta_R(D) + {1\over 2}f_D(z/L)\right]
\eeq
Notice again that in these expressions we have summed over the transverse indices $i$ and $j$, each taking $D-2$ different values. All these correlators
are seen to diverge near the plates $z\ra 0$ or $z\ra L$ where the function $f_D(z/L)$ diverges. This is the same phenomenon which has previously been
seen in $D=4$ dimensions\cite{LR}.

The pressure betwen the plates due to these fluctuations is defined by $P = \ex{T_{zz}}$. From (\ref{T_EM}) we have $T_{zz} = B_{iz}^2 - E_z^2 + {\cal L}$ 
where now
\beq
            \ex{{\cal L}} = -{1\over 2}(D-1){(D-2)\Gamma(D/2)\over (4\pi)^{D/2}L^D}f_D(z/L)
\eeq
Together with the values for $\ex{E_z^2}$ and $\ex{B_{iz}^2}$ from above, the $z$-dependence from the function $f_D(z/L)$ cancels out in the pressure 
and gives the expected value (\ref{press}). 

So far there are no inconsistencies in the obtained results. But when we now calculate the energy density ${\cal E} = \ex{T_{00}}$ between the plates,
with $T_{00} = E_i^2 + E_z^2 - {\cal L}$, we obtain
\beq
          {\cal E} = - {(D-2)\Gamma(D/2)\over (4\pi)^{D/2}L^D}\left[\zeta_R(D) - (D/2 - 2)f_D(z/L)\right]         \label{E_dens}
\eeq
The $z$-dependence in the last term is non-zero when $D>4$ and makes the energy density diverge like $z^{-D}$ with distance $z$ from the plates. As a result, the total energy of the system is
infinite, a result which seems to be impossible to reconcile with the finite Casimir force (\ref{press}). In fact, (\ref{press})  corresponds to having a 
constant energy density equal to the first term in (\ref{E_dens}). This apparent inconsistency has been verified in a different approach based on Green's 
function methods\cite{AORW}.

It is tempting to explain this problem by the imposed boundary conditions. We have used the MIT boundary condition which is equivalent to letting
the electromagnetic vector potential satisfy Neumann boundary conditions in the axial gauge. Had we instead imposed metallic boundary conditions, equivalent to Dirichlet boundary conditions for the 
vector potential in the axial gauge, the only change in the above results would be the replacement of the mode functions $\cos(n\pi z/L)$ with
$\sin(n\pi z/L)$ in (\ref{mode-ex}) so that $f_D \ra -f_D$ in the above results. Needless to say, the problem would remain. Only for periodic boundary 
conditions, as for finite temperature, would the disturbing term be absent\cite{ARW}. But this is not necessarily satisfying from a physical point of view. 
A more mathematical discussion of such divergences near confining boundaries has been initiated by Fulling but here only scalar fields are 
considered\cite{Fulling}.

A physical explanation of the above conumdrum becomes apparent when we take the limit $L\ra \infty$ and thus consider the quantum fluctuations around
a single plate. From (\ref{E_dens}) we then find the energy density
\beq
         {\cal E}_1 = (D-2)(D/2-2)\frac{\Gamma(D/2)}{(4\pi)^{D/2}{\vert z\vert}^D}              \label{singel}
\eeq
which is non-zero on both sides of the plate and diverges when we approach it. This situation is analogous to the diverging energy density surrounding 
a pointlike electron. It is intrinsic to a single plate and should not contribute to the interaction between the plates induced by the same 
vacuum fluctuations. To see the connection with the Casimir force, we should  subtract the self-energy (\ref{singel}) for both plates from the full energy 
density (\ref{E_dens}), taking into account both sides of each plate. We thus obtain the interaction energy density
\beq\!\!\!\!\!\!\!\!\!\!\!\!\!\!\!\!\!\!\!\!\!\!\!\!\!\!\!\!\!\!
   \tilde{\cal E} = -(D\!-\!2)\frac{\Gamma(D/2)}{(4\pi)^{D/2}L^D}\times\!
   \left\{
   \begin{array}{ll}
     (D/2 - 2)(L/(L-z))^D&\mbox{for $z < 0$,}\\
     \zeta_R(D) - (D/2 - 2)\tilde{f}_D(z/L)&\mbox{for $0 < z < L$,}\\
     (D/2 - 2)(L/z)^D&\mbox{for $z > L$,}
   \end{array}
   \right.
\label{renorm}
\eeq
where now

\beq
       \tilde{f}_D(z/L)=\zeta_H(D,1+z/L) +  \zeta_H(D,2 - z/L)
\eeq
It is seen to be finite everywhere, even at the plates. When integrating over the full volume, the $z$-dependent terms cancels out as follows from
\beq
       && \int_{-\infty}^0\!{dx\over(1-x)^D} - \int_0^1\!dx \tilde{f}(x) + \int_1^\infty\!{dx\over x^D}   =     \nn   \\
       && {2\over D-1}\left[1 + \sum_{n=0}^\infty \left( {1\over (n+2)^{D-1}} - {1\over (n+1)^{D-1}} \right)\right]=0
\eeq
Only the $z$-independent term in (\ref{renorm}) contributes and  agrees perfectly with the total energy corresponding to the Casimir force.
 
A similar and somewhat simpler system is the Casimir energy induced by a massless scalar field in the same geometry. One will then find a very 
similar result for the energy density as obtained here\cite{AORW}. It diverges near the plates for all spacetime dimensions $D >2$. Again this can 
be attributed to a divergent self-energy of each plate. However, when $D=2$ there are no such divergences and zero self-energy. But this is also the 
dimension in which the scalar theory has conformal invariance. In higher dimensions $D>2$ it is possible to make the scalar theory retain this invariance by 
adding a conformal term. The resulting, improved energy-momentum tensor\cite{Callan} then contains an additional piece discovered by Huggins\cite{Huggins} 
and makes it traceless. Including the Huggins term, the divergent part of the energy density corresponding to the last term in (\ref{E_dens}) drops out 
as s first noticed by de Witt when $D=4$\cite{dW}.

For the electromagnetic field we have used the canonical energy-momentum tensor (\ref{T_EM}) which has the trace $T^\mu_{\;\;\,\mu} =(4-D){\cal L}$. It is
zero for $D=4$ which reflects the well-known fact that Maxwell theory is then conformally invariant. There are then no diverences in the
Casimir energy. Thus it is natural to relate the apparent inconsistency in the electromagnetic Casimir energy when $D>4$  to the lack of conformal invariance. 
It does not seem to be possible to construct an improved energy-momentum tensor in this case because gauge invariance forbids the existence 
of any corresponding local Huggins term. From this point of view the divergent, electromagnetic self-energy can therefore not be removed. For this to be done, 
one needs a more realistic description of the boundary plates along the lines considered by others\cite{phys_bc}.

This work has been supported by the grants NFR 159637/V30 and NFR 151574/V30 from the Research Council of Norway.

\end{document}